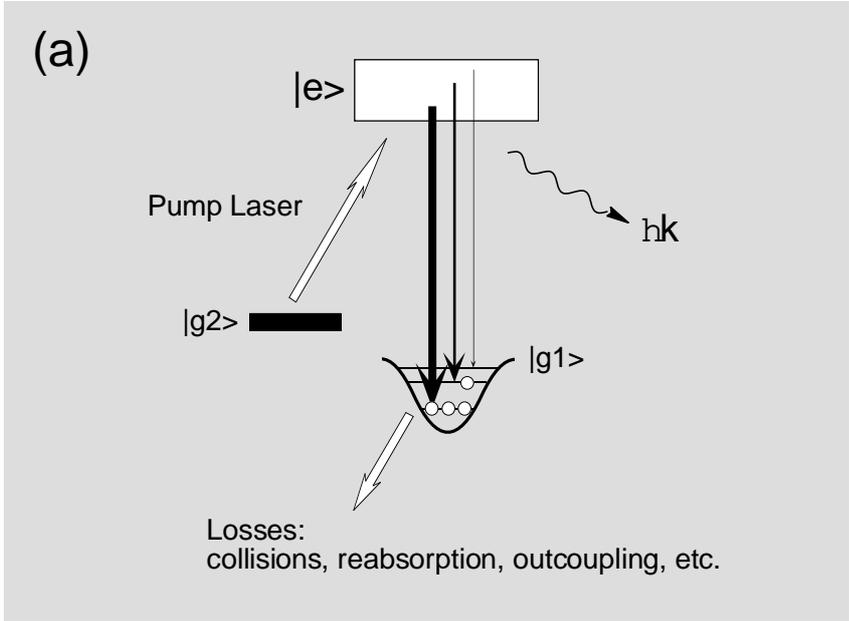

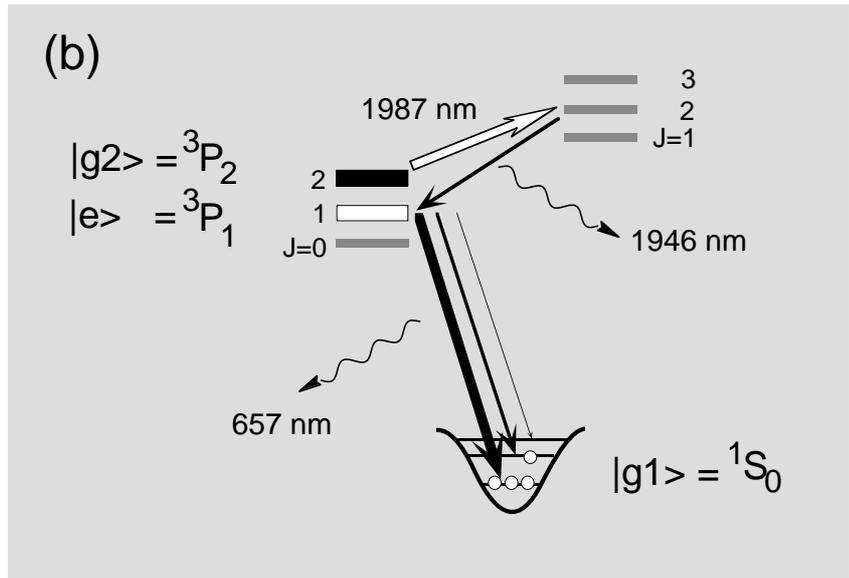

J. Grünert et al., fig.1

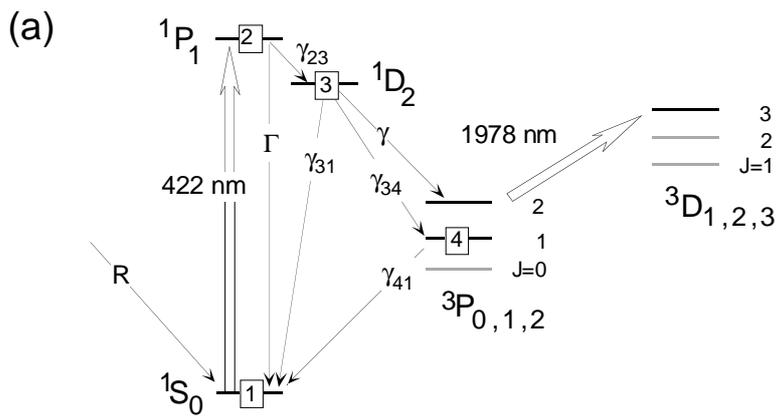

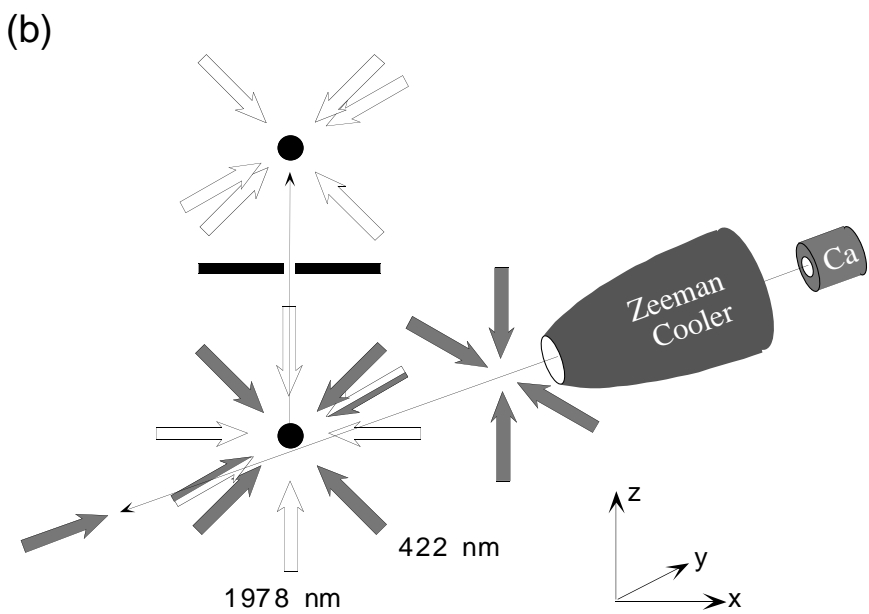

J. Grünert et al., fig.2

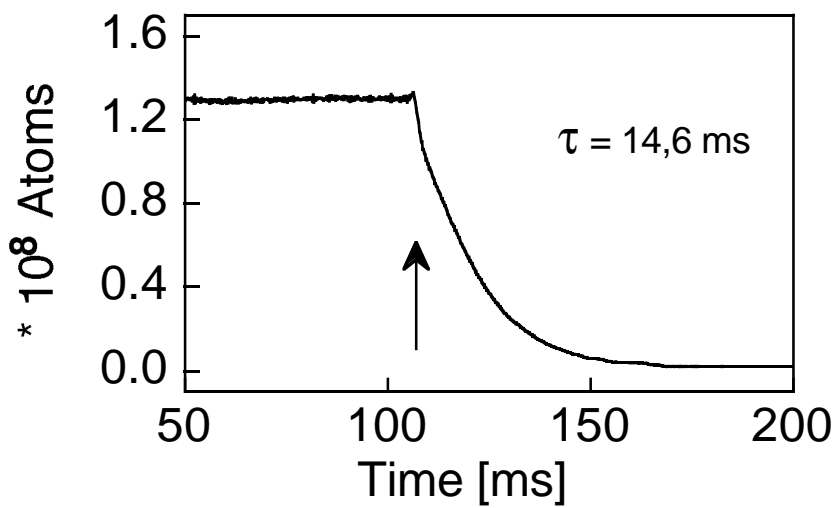

J. Grünert et al., fig.3

# Ultracold Metastable Calcium Ensembles, a Medium for Matter Wave Amplification ?

J. Grünert, G. Quehl, V. Elman, and A. Hemmerich

*Institut für Laserphysik, Universität Hamburg, Jungiusstraße 9, D-20355 Hamburg, Germany*

We propose an experimental implementation of matter wave amplification by optical pumping (MAO) with metastable calcium atoms. First experimental results indicate that pumping rates can be significantly higher than in previous experimental schemes and let it appear promising that the threshold condition for generation of degeneracy can be reached.

   The effect of stimulated emission of photons in a laser is based on their bosonic nature. Consequently, an analogous phenomenon should occur for other bosonic particles as for example atoms with integer total angular momentum. In fact, in the late nineties, following the first experimental observation of Bose-Einstein condensation (BEC) [1], schemes for matter wave amplification have begun to attract interest and theoretical models for matter wave analogs of the laser have been proposed, which employ the mechanism of optical pumping as a key element [2,3]. Matter wave amplification by optical pumping (MAO) might open a novel path for a continous production of coherent matter waves at rates exceeding those accessible via BEC by orders of magnitude.

   A fundamental prerequisite for MAO are atoms which provide a so called Λ-level scheme comprised of two stable, energetically well separated ground states, |g1> and |g2>, and one excited state |e>, that allows for controlled optical pumping of atomic population from |g2> to |g1> as sketched in fig.1(a). By means of laser cooling techniques a dense cold sample of atoms is continuously produced in the reservoir state |g2> which plays the role of the inverted medium in a laser. Atoms in |g1> are captured inside a potential well resulting for example from a far-detuned optical dipole trap. These atoms play the role of the photons stored in some cavity mode in a laser. Optical pumping, i.e. excitation to |e> and subsequent spontaneous emission can transfer atoms from |g2> into |g1>. Due to the bosonic character of the atoms the probability to end up in the motional state | g1,ν> scales with the number $N_\nu$ of atoms already populating | g1,ν>. This is analogous to the effect of stimulated emission in a laser. If the optical pumping rate for some motional state | g1,ν> exceed the losses, a macroscopic population of this state should arise. The motional ground state | g1, 0> typically experiences the highest pumping and the lowest loss rate and should thus exhibit the lowest threshold for degeneracy.

   Unfortunately, in traps with an extension much larger than the wavelength of the optical pumping photons with respect to all three dimensions it is argued that the threshold condition for the onset of degeneracy cannot be reached due to the possible reabsorption of the optical pumping photons which heat the trapped atomic sample [4]. Traps, which are microscopic in three dimensions reach threshold at such high densities that three-body collisions may yield formation of molecules, and large pumping rates are difficult to achieve due to the small trap volume. One way out of this dilemma may be to surpass the threshold in a microscopic trap and subsequently to perform an adiabatic increase of the trap volume [5]. A simpler way around the reabsorption problem is to employ strongly asymmetric traps from which the optical pumping photons can escape with a high probability [3]. We can roughly estimate under which conditions threshold should be reached despite of reabsorption, for example, in the case of a cylindrical trap of radius r and length l. The probability for an atom in |e> within the trap volume V= $\pi r^2 l$ (with motional degrees of freedom described by a thermal state with temperature T and thermal de Broglie wavelength $\lambda_{th}$) to be transferred to the motional ground state of |g1> in an optical pumping process, is approximately given by a Franck-Condon factor which can be estimated to be $P_{gain} = \lambda_{th}^3/V$ [6]. This has to be compared with the reabsorption probability which is approximated by $P_{loss} = \sigma/l^2$, where $\sigma = \lambda_p^2/2\pi$ denotes the resonant absorption cross section for the optical pumping photons of wavelength $\lambda_p$. Thus, $P_{gain} \geq P_{loss}$ is equivalent to $2 * (\lambda_{th}/\lambda_p)^2 * (\lambda_{th}/r) * (l/r) \geq 1$ which can be satisfied for the realistic combination of parameters $(\lambda_{th}/\lambda_p) = 10^{-1}$, $(\lambda_{th}/r) = 5*10^{-2}$ and $(l/r) = 1000$. With this trap geometry in mind we neglect the reabsorption problem in order to roughly estimate the condition for the onset of degeneracy, finding that the pump rate per trap volume V into state |e> should exceed $R_0 \approx \kappa V/\lambda_{th}^3$ where $\kappa$ is the ground state loss rate [6]. At 1 microkelvin $\lambda_{th}$ is typically on the order of 50 nm and loss rates can be as low as $10^{-2}$ s$^{-1}$ yielding $R_0/V = 8*10^{13}$ cm$^{-3}$s$^{-1}$. For the cylindric trap considered above this corresponds to $R_0 = 2,5*10^7$ atoms/s into the trap volume. To realize such high transfer rates of atoms at such low temperatures presents a serious challenge for the experimentalist.

In this article we discuss an experimental implementation of atomic matter wave amplification by optical pumping based on ultracold metastable calcium atoms. Calcium owing to its two valence electrons provides a singlet and a triplet ground state, well separated energetically, which possess sufficiently long life times in order to serve as the states |g1> and |g2> respectively (see fig.1(b)). The extremely long life time of the $^3P_2$ triplet state (the reservoir state |g2>) results because the transition to the singlet ground state violates two selection rules concerning spin and total angular momentum. An infrared laser of approximately 2 μm wavelength can optically pump atoms from $^3P_2$ via $^3D_2$ to $^3P_1$, which has a sufficiently short life time of approximately 0,4 ms and thus may serve as the excited state |e> in fig.1(b). The transition at 657 nm connecting $^3P_1$ to the singlet ground state $^1S_0$ (which serves as the trapped state |g1>) is the famous red intercombination line. The singlet ground state has zero total angular momentum and possesses no further degeneracy. It is basically insensitive against magnetic fields but can be trapped in a far-detuned dipole trap. As a notable advantage over similar schemes employing noble gases (e.g. argon [7]) which also provide two stable electronic states, in our scheme the |g1> atoms are real ground state atoms such that no heating collisions (e.g. hyperfine structure changing collisions, Penning ionization etc.) can arise.

The experimental task is to produce cold calcium atoms at 1 μK in the $^3P_2$ triplet state at a rate in the $10^{10}$ s$^{-1}$ range. Figure 2 indicates how this goal can be achieved in two steps by laser cooling. Calcium possesses an extremely efficient Doppler-cooling line at 422 nm (natural linewidth $\Gamma/2\pi = 34,6$ MHz) connecting the singlet ground state $^1S_0$ to the singlet $^1P_1$ state. This can be exploited to load a magneto-optic trap (MOT) from a Zeeman-cooled atomic beam. Capture rates above $10^{11}$ atoms/s should present no principle difficulty. As is indicated in fig.2(a), the $^1S_0 \rightarrow {}^1P_1$ transition is not a perfectly closed transition, a feature shared by all earth alkaline atoms with the exception of magnesium. The excited $^1P_1$ singlet level may decay into the lower lying $^1D_2$ state with a rate of approximately $\gamma_1 = 2180$ sec$^{-1}$. About 78 % of these atoms return to the ground state in approximately 3 ms either directly in a quadrupol transition or via the $^3P_1$ state. These atoms are not captured during this time and thus the capture volume of the MOT has to be sufficiently large (> 1 cm) to recycle them when they return to the ground state. The remaining 22 % are transferred to the metastable $^3P_2$ state and cannot be recaptured at all. This is the dominant trap loss mechanism. For the case of perfect recycling the production rate of $^3P_2$ atoms equals the capture rate of the MOT and thus can exceed $10^{11}$ atoms/s.

The $^3P_2$ atoms are produced at a temperature given by the Doppler cooling limit for the 422 nm cooling line of 0,8 mK. This is sufficiently low to capture them in a second MOT employing the closed $^3P_2 \to {}^3D_3$ transition at 1978 nm. The narrow linewidth of this transition of about 57 kHz together with its long wavelength provides a very low Doppler-limit of only 1,3 µK and a recoil limit of 122 nK. Moreover, the J=2→J=3 level structure allows for polarization gradient cooling, such that temperatures around a microkelvin should be obtained. For implementation of the MAO scheme we can not tolerate the presence of 422 nm photons which would excite atoms to $^1P_1$ (cf. fig.1(b)). Thus, a cold beam has to be extracted from the 1978 nm MOT into a second vacuum chamber (with background pressure below $10^{-11}$ mBar) and recaptured in a second MOT at 1978 nm (cf. fig. 2(b)). This has been established as a standard procedure in laser cooling experiments with alkalis and should not yield any difficulties like loss of atoms. Note that the 1978 nm MOT is compatible with the synchronous operation of a dipole trap for the singlet ground state.

While for alkalis magneto-optic trapping has become a standard technique that has been established in many laboratories, yet only a few research groups have operated such traps for earth alkaline elements. Traps for strontium [8,9,10], calcium [8,11,12] and magnesium [13] have been reported with numbers of trapped atoms in the $10^7$ range. In contrast to experiments with alkalis which are readily addressed by semiconductor lasers and which can be captured from room temperature vapors, experiments with earth alkaline atoms have involved highly complex atomic beam machines and laser systems. We have built a compact and efficient experimental setup consisting of a miniature vacuum apparatus and all solid state light sources. Our experiment is sketched in fig. 2(b). Calcium atoms are evaporated at 600°C through 45 adjacent channels of 10 mm length and 1 mm diameter. The atoms are decelerated by a laser beam counterpropagating the atoms inside a 30 cm long tapered solenoid. A spatially varying magnetic field compensates the Doppler shift of the decelerated atoms. After the slowing magnet an additional coil operated with reversed direction of current serves to produce a relative minimum of the magnetic field in order to decouple the atoms from the slowing process at a defined velocity. During this decoupling process the atoms are cooled by a two-dimensional optical molasses along the transverse directions. Subsequently, the cold atoms enter a small vacuum chamber, where three retro-reflected laser beams (Ø = 10 mm) are superposed with a quadrupole magnetic field in order to form a magneto-optic trap. The trapping region is located at 50 cm distance from the apertures of the source and 12 cm down stream from the transverse optical molasses.

The necessary 422 nm photons are made with an all solid state laser system starting off with a 5 W commercial frequency-doubled YAG system pumping a miniaturized Ti:Sapphire laser which produces 800 mW dual frequency output at 845 nm [14]. The two frequencies separated by about 1 GHz are used for sum frequency mixing in a $KNbO_3$ crystal. yielding a total blue output of 150 mW consisting of 3 frequency components. The sum frequency component (approximately 100 mW) is used for the experiment. About 70 mW is directed the Zeeman slower and the transverse molasses while 30 mW are used for the MOT beams, which has been found to yield optimized results. The available 422 nm power is limited by thermal lenses typical for $KNbO_3$ already at modest fundamental powers. Increasing the available fundamental power to 2 W and using LBO for sum frequency mixing should lead to a factor five increase of 422 nm output. In order to generate narrow band radiation at 1978 nm we have developed a miniaturized, Tm:YAG laser based on the same design criteria as used for our Ti:Sapphire laser. We obtain up to 90 mW of tunable dual frequency radiation [15], far more than needed for our infrared MOT.

For interpretation of our experimental findings we employ a rate equation model including the levels $^1S_0$, $^1P_1$, $^1D_2$, $^3P_1$, indicated in this order by 1,2,3,4 in fig. 2(a). Neglecting collisional loss due to hot background atoms the rate equations are:

$$\begin{aligned}
\dot{N}_1 &= R - W N_1 + (\Gamma + W) N_2 + \eta \gamma_{31} N_3 + \eta \gamma_{41} N_4 \\
\dot{N}_2 &= W N_1 - (\Gamma + \gamma_{23} + W) N_2 \\
\dot{N}_3 &= \gamma_{23} N_2 - (\gamma_{34} + \gamma_{31} + \gamma) N_3 \\
\dot{N}_4 &= \gamma_{34} N_3 - \gamma_{41} N_4
\end{aligned} \quad (1)$$

Here $N_i$ is the number of trapped atoms in state i, $\eta$ is the fraction of those atoms which after having decayed back to the ground state can be recaptured while R and W denote the 422 nm MOT-capture rate and the excitation rate for the $^1S_0 \to {}^1P_1$ transition. The steady state solution of eq.1 yields the production rate $\hat{R} = N_3 \gamma$ of $^3P_2$ atoms to be $\hat{R} = \varepsilon R$, where R denotes the 422 nm MOT capture rate and $\varepsilon = \gamma [(1-\eta)(\gamma_{34} + \gamma_{31}) + \gamma]^{-1}$ varies between 0,22 and 1 depending on whether there is no ($\eta=0$) or perfect ($\eta=1$) recycling. If $\Gamma \gg W$ and $\gamma_{41} \gg \gamma_{34}$, the population $N_2$ evolves much more rapidly than $N_1$ and thus takes its steady state value according to the temporary value of $N_1$. Thus we may approximate $N_2/N_1 \approx W/(\Gamma + \gamma_{23} + W)$. Similarly, if $\gamma_{41} \gg \gamma_{34}$, we may approximate $N_4/N_3 \approx \gamma_{34}/\gamma_{41}$. With these approximations we obtain reduced rate equations

$$\frac{\partial}{\partial t}\begin{pmatrix} N_2 \\ N_3 \end{pmatrix} = \begin{bmatrix} -\sigma\gamma_{23} & (\gamma_{31}+\gamma_{34})\sigma\eta \\ \gamma_{23} & -\gamma_{31}-\gamma_{34}-\gamma \end{bmatrix} \begin{pmatrix} N_2 \\ N_3 \end{pmatrix} + \begin{pmatrix} \sigma R \\ 0 \end{pmatrix}, \quad (2)$$

where $\sigma = s/(2+s)$ and s is the saturation parameter of the $^1S_0 \to {}^1P_1$ transition. Setting R=0 in eq.2 we expect a biexponential decay of the populations $N_2$ and $N_3$ with decay rates given by the eigenvalues of the right hand side matrix which are approximately given by

$$\gamma_\pm = \frac{1}{2}(\sigma\gamma_{23}+\gamma_{31}+\gamma_{34}+\gamma) \pm \sqrt{\gamma_{23}(\gamma_{31}+\gamma_{34})\sigma\eta} \quad (3)$$

if the matrix elements a,b,c,d satisfy $(a-d)^2 \ll bc$. Using eq.3 we may obtain a value for $\eta$ and thus a value for $\varepsilon$ by measuring the dominant component $\gamma_-$ in the decay of $N_2$. This quantity is directly accessible in our experiment via observation of the 422 nm fluorescence of the decaying MOT after shutting off the atomic beam cooling. The observed decay within about 15 ms (cf. fig.3) lets us estimate the recycling efficiency $\eta$ to be 91 %, the corresponding value of $\varepsilon$ is 0,76. Here we have assumed s = 0,5 which corresponds to the laser power and trapping beam diameters in our MOT. The rate $\hat{R}$ can be directly connected to the number $N(^1P_1)$ of atoms in the $^1P_1$-state by $\hat{R} = N(^1P_1) * \gamma_{23} \gamma (\gamma_{31} + \gamma_{34} + \gamma)^{-1} = N(^1P_1) * 480$ s$^{-1}$. The value of $N(^1P_1)$ is given by $P/\hbar\omega\Gamma$, where P is the total 422 nm fluorescence power and $\hbar\omega$ is the energy per 422 nm photon. In our preliminary experiments we find that $N(^1P_1) = 2,2*10^7$ and correspondingly $\hat{R} = 1,1*10^{10}$ s$^{-1}$. Assuming again s = 0,5 we have $1,4*10^8$ calcium atoms in the $^1S_0$ ground state. This represents the highest number of trapped calcium atoms reported so far, although important components of our experiment are not yet optimized. The diameters of our trapped atomic samples are typically 5 mm corresponding to a density of about $2,1*10^9$ atoms/cm$^3$. The sizes of the trapped samples decrease with decreasing loading rate, showing that our MOT operates in the density limited regime [16].

Similarly as in ref.[12], we have used 672 nm light in order to repump 3 $^1D_2$-atoms via 5 $^1P_1$ back to the singlet ground state 4 $^1S_0$ and thus close the dominant loss channel of our MOT. First experiments show only a factor 3 to 4 increase of 422 nm fluorescence in presence of 672 nm light, in contrast to the expectations of a factor of about 100 if the residual trap loss would result from collisions with hot background atoms. This may be partly due to incomplete repumping because of inhomogeneous magnetic broadening and the existence of non-coupling states for the repumping transition. Moreover, one may speculate that, similarly as shown for strontium in [10], collisions between cold 4 $^1S_0$ and 4 $^1P_1$ atoms may yield an additional loss channel owing to the effect of radiative redistribution [17] which can be particularly efficient due to the existence of a metastable $^1\Pi_g$ molecular state for the calcium dimer.

Our experimental observations make it appear promising to surpass the threshold for quantum degeneracy in future experiments as is seen from the following estimation. Under the realistic assumption that we can in fact cool all mestastable atoms down to 1 μK into a volume of $V_{cool}$ = 1 mm$^3$, we may estimate the number of atoms pumped per second into our example cylinder trap (radius r = 1 μm, length l = 1 mm, volume V= πr$^2$l ) to be $\widetilde{R}_0 = \gamma \widehat{R} V/V_{cool}$ where γ = 2,6∗10$^3$ s$^{-1}$ is the natural decay rate of the $^3P_1$ triplet state. This yields $\widetilde{R}_0$ = 9 ∗10$^7$ s$^{-1}$ which is to be compared with the theoretical threshold rate $R_0$ = 2,5∗10$^7$ s$^{-1}$. Collisions between cold metastable atoms, particularly in the presence of infrared photons, might complicate the situation providing a challenge of future theoretical work to investigate these.

In summary, we have proposed a novel experimental implementation for matter wave amplification by optical pumping employing laser cooled metastable calcium atoms. A bichromatic magneto-optic trap is under construction that operates in two successive cooling steps. Preliminary experiments show more than 10$^8$ trapped ground state calcium atoms and a production rate of more than 10$^{10}$ metastable calcium atoms per second. These results make it look promising that the threshold condition for the generation of quantum degeneracy can be approached.

We are grateful to A. Diening and G. Huber for providing us with a Tm:YAG crystal and valuable expertise. We thank F. Renzoni for constructive critical remarks. This work has been supported by the Deutsche Forschungsgemeinschaft under contract numbers DFG-He2334/2.1 and DFG-He2334/2.3.

**Figure Captions**

**Fig.1**
(a). Scheme for matter wave amplification by optical pumping (cf. ref.2).
(b). Our implementation of the scheme in (a) with calcium atoms. The transition rate for optical pumping metastable atoms into the singlet ground state is limited by the $^3P_1$ life time of 0,4 ms.

**Fig.2**
(a). Scheme for a bichromatic magneto-optic trap. The atoms, loaded from a laser cooled atomic beam at a rate R, are precooled by the strong Doppler-cooling line at 422 nm. Through leakage from the excited $^1P_1$ state to $^1D_2$ the metastable state $^3P_2$ is populated. Cooling via the narrow band infrared transition at 1978 nm should yield temperatures around 1 microkelvin. Atoms being transferred to $^1D_2$ are no longer trapped. About 78% of these atoms return to the ground state in typically 3 ms. These atoms can be recycled by providing a sufficiently large capture volume of the MOT. The levels labeled with numbers in square boxes are accounted for in eq.1.

(b). Schematic of experimental apparatus. The atomic beam enters the MOT region under an angle of 30° with respect to the z-axis and slightly displaced with respect to the center of the MOT in order to keep the hot atoms away from the cold ones.

**Fig.3**

The number of trapped atoms is plotted versus observation time. At the time indicated by the arrow, the Zeeman slowing laser beam is turned off. The decay of the trapped sample with a time constant of 14,6 ms is observed.

**References**

1. M. H. Anderson et al., Science **269**, 198 (1995). K. Davis et al., Phys. Rev. Lett. **75**, 3969 (1995).
2. R. Spreeuw, T. Pfau, U. Janicke, and M. Wilkens, Europhys. Lett **32**, 469 (1995).
3. M. Olshanii, Y. Castin, and J. Dalibard, in A. Sasso, M. Inguscio, M. Allegrini, eds., Procs. of the XII Conference on laser spectroscopy, world scientific, New York (1995).
4. U. Janicke and M. Wilkens, Europhys. Lett. **35**, 561 (1996).
5. I. Cirac and M. Lewenstein, Phys. Rev. A **53**, 2466 (1996). I. Castin, J. Cirac, and M. Lewenstein, Phys. Rev. Lett. **80**, 5305 (1998).
6. U. Janicke and M. Wilkens, Adv. Mol. Opt. Phys. **4** (1999).
7. H. Gauck, M. Hartl, D. Schneble, H. Schnitzler, T. Pfau, and J. Mlynek, Phys. Rev. Lett.**81**, 5298 (1998).
8. T. Kurosu and F. Shimizu, Jpn. J. Appl. Phys. **29**, L2127-L2129 (1992).
9. H. Katori, T. Ido, Y. Isoya, and M. Kuwata-Gonokami, Phys. Rev. Lett. **82**, 1116 (1999).
10. T. P. Dineen, K. R. Vogel, E. Arimondo, J. L. Hall, A. Gallagher, Phys. Rev. A **59**, 1216 (1999).
11. T. Kisters, K. Zeiske, F. Riehle, and J. Helmcke, Appl. Phys. B **59**, 89 (1994).
12. C. W. Oates, F. Bondu, R. W. Fox, and L. Hollberg, Eur.Phys.J. D **7**, 449 (1999).
13. F. Ruschewitz, J. L. Peng, H. Hinderthür, N. Schaffrath, K. Sengstock, and W. Ertmer, Phys. Rev. Lett. **80**, 3173 (1998).
14. C. Zimmermann, V. Vuletic, A. Hemmerich, L. Ricci, T. Hänsch, Opt. Lett. **20**, 297 (1995).
15. G. Quehl, J. Grünert, V. Elman, and A. Hemmerich, to be published (2000).
16. C. G. Townsend, et al., Phys. Rev. A **52**, 1423 (1995).
17. A. Gallagher and D. E. Pritchard, Phys. Rev. Lett. **63**, 957 (1989).